# Automating Systematic Literature Reviews with Natural Language Processing and Text Mining: a Systematic Literature Review

Girish Sundaram[1] and Daniel Berleant[1]

[1] University of Arkansas at Little Rock, Little Rock AR 72204, USA
gsundaram@ualr.edu, jdberleant@ualr.edu

**Abstract.**

Objectives: An SLR is presented focusing on text mining based automation of SLR creation. The present review identifies the objectives of the automation studies and the aspects of those steps that were automated. In so doing, the various ML techniques used, challenges, limitations and scope of further research are explained.

Methods: Accessible published literature studies that primarily focus on automation of study selection, study quality assessment, data extraction and data synthesis portions of SLR. Twenty-nine studies were analyzed.

Results: This review identifies the objectives of the automation studies, steps within the study selection, study quality assessment, data extraction and data synthesis portions that were automated, the various ML techniques used, challenges, limitations and scope of further research.

Discussion: We describe uses of NLP/TM techniques to support increased automation of systematic literature reviews. This area has attracted increase attention in the last decade due to significant gaps in the applicability of TM to automate steps in the SLR process. There are significant gaps in the application of TM and related automation techniques in the areas of data extraction, monitoring, quality assessment and data synthesis. There is thus a need for continued progress in this area, and this is expected to ultimately significantly facilitate the construction of systematic literature reviews.

## 1 Background

In this section we describe the motivation of our work beginning with a brief overview of Systematic Literature Reviews (SLRs) especially about the study selection, study quality assessment, data extraction and data synthesis phases within SLRs. We then also review the existing prior arts to get more insights on the gaps that exist in this field.

A systematic review is one of the numerous types of reviews [1] and is defined as [2] "a review of the evidence on a clearly formulated question that uses systematic and explicit methods to identify, select and critically appraise relevant primary research, and to extract and analyze data from the studies that are included in the review." The methods used must be reproducible and transparent.

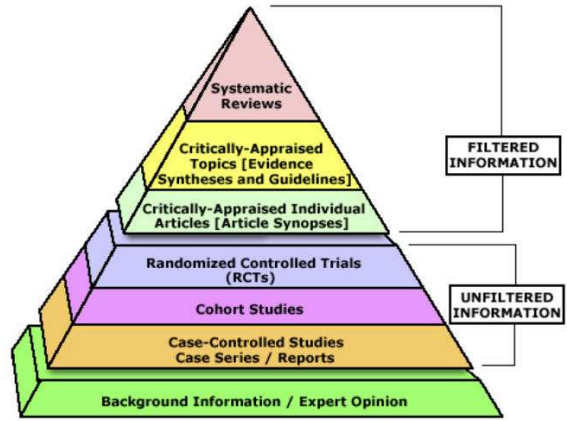

**Figure 1**. Systematic Reviews (Based on Glover et al. 2006) [3]

Fig. 1 illustrates that systematic reviews are considered to provide the highest quality of evidence in the area of evidence based medicine (EBM). While SLRs are the norm in the field of EBM and healthcare, Kitchenham and Charters [4] provided the framework and guidelines on how SLRs can be used in other fields like software engineering.

Preparing an SLR can be both time consuming and expensive [5-6]. The time problem is further accentuated by the fact that SLRs become outdated, making their timely completion a quality factor. Shojania et al. [7] shows that the median lifetime of an existing review until it needs updating is 5.5 years. It is apparent that the current SLR process needs augmentation to speed up the process of creating them.

Specific phases of SLR development such as identification of relevant studies, data collection, extraction and synthesis have been found to require time consuming and error prone manual effort [8].

NLP and text mining have been used increasingly in the recent past to analyze and automate steps in the SLR process. This paper performs an SLR on the current state of the art. One objective of performing this SLR is to identify specific steps where there has been considerable activity and where there is a scope for further research. We have adapted Table 1 from Kitchenham and Charters [4] to name the steps within the SLR process.

Our primary steps of interest as part of this SLR are SLR5 – SLR9. Studies have shown that steps SLR6 – SLR9 are often among the most time consuming [9-11].

**Table 1.** Key Steps in Systematic Literature Reviews
(Based on Kitchenham & Charters [4])

| ID | Category | Step | Synonyms |
|---|---|---|---|
| SLR1 | **Defining a review** | Commissioning a review | |
| SLR2 | | Defining the research questions | |
| SLR3 | | Determining a protocol for the review | |
| SLR4 | | Evaluating the protocol for the review | |
| SLR5 | **Conducting the review** | Identification of research | Literature search, search string development |
| SLR6 | | Selection of studies | Citation screening |
| SLR7 | | Assessing study quality | Selection review |
| SLR8 | | Data extraction & monitoring | |
| SLR9 | | Data synthesis | |
| SLR10 | **Reporting the review** | Specifying dissemination mechanisms | |
| SLR11 | | Formatting the main report | |
| SLR12 | | Evaluating the report | |

## 2 Summary of Previous Reviews

Here we describe other reviews of SLR automation to help place the present study in context. Table 2 lists them briefly, followed by additional details.

**Table 2.** Related Studies

| No. | Title | Objective | Reference |
|---|---|---|---|
| 1. | Automating data extraction in systematic reviews: a systematic review | A systematic review focusing on automatic data extraction prior arts. | Jonnalagadda et al. [12] |
| 2. | Using text mining for study identification in systematic reviews: a systematic review of current approaches | A systematic review focusing on identifying relevant articles using the title and abstract for reference. | O'Mara-Eves et al. [13] |
| 3. | Text-mining techniques and tools for systematic | Explains how text mining techniques can contribute to SLR development, focusing on the following text mining | Feng et al. [14] |

| | | |
|---|---|---|
| | literature reviews: a systematic literature review | categories, namely, information extraction, information retrieval, information visualization, classification, clustering and summarization. | |
| **4.** | Systematic review automation technologies | A systematic review to study the feasibility of automating various phases in an SLR. | Tsafnat et al. [15] |
| **5.** | Toward systematic review automation: a practical guide to using machine learning tools in research synthesis | A guide that can be used by SLR researchers to apply machine learning methods to reduce the overall turnaround time. | Marshall et al. [16] |
| **6.** | Moving toward the automation of the systematic review process: a summary of discussions at the second meeting of the International Collaboration for the Automation of Systematic Reviews | Documents various ongoing short term projects that are carrying out research in the automation of SLR. | O'Connor et al. [17] |
| **7.** | Making progress with the automation of systematic reviews: principles of the International Collaboration for the Automation of Systematic Reviews | Documents the outcomes from the conference to improve the overall efficiency of conducting a SLR. | Beller et al. [18] |
| **8.** | Usage of automation tools in systematic reviews | Documents the potential issues that reviewers face when trying to use SLR automation techniques. | Van Altena et al. [19] |
| **9.** | A critical analysis of studies that address the use of text mining for citation screening in systematic reviews | Reviews text mining in the context of systematic literature reviews. More specifically, the focus is on one task within the systematic literature review process. That task is screening citations to determine which ones to include in the review. | Olorisade et al. [20] |

Jonnalagadda et al. [12] focus on automatic data extraction of critical data elements from full text medical texts as part of the SLR process. They identified 52 potential data elements used in systematic reviews which the authors obtained from standard medical databases/tools such as Cochrane Handbook for Systematic Reviews [21], The CON-solidated Standards Of Reporting Trials (CONSORT) statement, The Standards for Reporting of Diagnostic Accuracy (STARD) initiative, PICO [22], PECODR [23] and PIBOSO [24]. The authors concluded that there is no unified data extraction framework that is focused on SLRs and the prior arts were limited in their scope of the number of data elements (1-7) that were considered in this step. NLP has been limited in its application in this field and there is considerable scope in further improving its involvement in the data extraction phase of SLRs.

O'Mara-Eves et al. [13] focus on the screening phase of SLR development which is time consuming and this is further accentuated by the rapid growth in the number of publications in the medical domain. Reviewers have to manually scan through a long list of mostly irrelevant articles that a search yields to identify relevant publications. The paper proposes a solution to semi-automate the screening phase of the SLR process. There are two processes in text mining that can help in screening. One is providing a prioritized list of items, with the ones on the top being the most relevant, that can be used for manual screening by a reviewer. The other is to use machine learning techniques where the system learns from a list of manual classifications of studies as included, or not, and then is able to automatically apply those classifications. They found that both the approaches resulted in reduction of the workload but it was not conclusively proven which method was superior. Some of the research also points out that the performance of the machine learning based system for relevant article prediction is similar to human efficiency. There is significant potential that this phase can be further improved to reduce the workload in the process [25].

Feng et al. [14] conducted an SLR to identify and classify text mining (TM) techniques to support the SLR process. They classified the various text mining techniques into 6 different categories: information extraction (IE), information retrieval (IR), information visualization (IVi), document classification and clustering and document summarization. As per their search methodology, out of the shortlisted papers, a majority were focusing on identifying the relevant articles for the study selection stage. They found that the four main applications of TM techniques are, 1) visual text mining (VTM), 2) federated search, 3) automated document/text classification and 4) document summarization. The researchers also attempted to answer the important question of which SLR activities could potentially benefit from TM. There is limited application of TM in the pre-review mapping study as part of the planning phase that Kitchenham & Charters [4] recommend. This is an important phase since the quality of an SLR is directly related to the protocol definition and scoping. There is scope for improvement in the query string development process which is the primary means to locate relevant studies from a variety of sources. Shortlisting and creating the finalized list of articles of interest is the phase where VTM techniques that combine clustering and information visualization have been used by many researchers.

Tsafnat et al. [15] surveyed the literature focusing on automating all aspects of SLRs and found that some of the tasks are fully automatable while many are not. They broke down the SLR process into 4 distinct steps, namely preparation, retrieval, appraisal, synthesis and write-up and examined the current level and future prospects of automation for these steps. Current research such as global evidence maps and scoping studies can guide in identifying gaps in the work done, helping to provide decision support for reviewers to fine tune and prioritize the research questions [26-28]. They found that despite available tools such as the Cochrane Database of Systematic Reviews and others, creating specific search filters to find relevant items is still a time consuming manual task. There an opportunity to create specialized systems that can understand the nuances of SLR questions and translate them into search filters for efficient identification of relevant prior work. Doing this currently requires specialized expertise in medicine, library science and standards. Templates such as the ones provided by Cochrane Review Manager [29] can be a good starting point and there is on-going research in this area.

Computational reasoning tasks have not been used extensively in SLRs [30-32]. The associated language bias problem helps define the scope of further research in the application of OCR and NLP to query definition. The application of Automatic Query Expansion (AQE) (synonym expansion, word sense disambiguation, auto-correction etc.) is part of the search phase of the SLR process. There is little existing work that automates and replicates sequential searching, removing duplicates or auto-screening of results, which experts use to progressively tune the search parameters based on relevancy of the search results. Related areas include automating snowballing (pursuing references of references) [33] and auto-extraction of important information such as trial features, methods, and outcomes from the texts of shortlisted literature. Overall the authors conclude that there are significant potential benefits of automating the SLR process using AI/ML.

Marshall et al. [16] review with practical examples how automation technologies can be used, situations where they might help, strengths/weaknesses, and how an SLR team can put these technologies into practice. Although significant work is being done in this area, concerns about accuracy of the current processes limit adoption and highlight the advantage of "human in the loop" automation rather than full automation. Search automation to expedite identification of relevant articles is the most advanced and commercial tools such as Abstrackr, RobotAnalyst, and EPPI reviewer have been used for secondary screening.

O'Connor et al. [17] and Beller et al. [18] discuss the proceedings of the Int. Collaboration for Automation of Systematic Reviews (ICASR). The authors observed that the number of data sets and tools for automation is increasing while at the same time integrating  the various tools into a workflow remains a challenge. Most of the tools for screening consider only the abstract and title for classification, but using the full text is complicated by the fact that most of the articles are in PDF format.  Acceptance of automation tools is limited due to concerns about accuracy and validity.

Van Altena et al. [19] write about the issues with adoption of automation tools in the SLR process. Candidate tools for their survey conducted were chosen from the "Systematic Review Toolbox" [34] website, which is a comprehensive list of tools compiled by researchers in the field. The survey results point out that researchers are willing to consider automation and generally feel that they can help. At the same time there are deterrent factors like poor usability, steep learning curves, lack of support and difficulty in integrating into a workflow. Another important observation was most of the tools did not have the ability to explain how the results were produced.

Olorisade et al. [20] critically analyzed the various text mining techniques used to augment the SLR process, specifically focusing on the citation screening phase. Certain models like support vector machines (SVM), Naïve Bayes (NB) and committee of classifiers ensembles were the most commonly used. Current automation research for SLR development focuses on study identification, citation screening and data extraction using tools such as SLuRp, StArt, SLR-Tool and SESRA [35-38]. They found missing or insufficient information about details needed for replicating research results such as number of support vectors being used in an SVM model or the number of neurons or hidden layers used. Progress seems slow given the amount of research being done in this area.

There is a need for SLRs focusing on automation of study selection, study quality assessment, data extraction and data synthesis using TM techniques including NLP, which TM techniques work the best, and performance and accuracy comparisons of human vs AI/TM driven approaches. Thus there is a need for an SLR which focused on these questions.

## 3 Identifying the Articles for This SLR

As we have seen in the related work section there is a need for SLRs that focus on SLR automation using TM and AI. The field of AI/TM is growing rapidly and such an SLR is expected to help accelerate further studies in this domain. We follow the guidelines explained by Kitchenham & Charters [4] which specifically highlight the need for a well-defined protocol to reduce bias, increase rigor and improve reproducibility.

### 3.1 Research Questions

From the software engineering prespective we aim to find all relevant information for the SLR process. The protocol for review defined by Gurbuz and Tekinerdogan [39] has been adapted here for this SLR.

Table 3 below documents the research questions to address. These questions are relevant from a text mining/ML perspective when building a model to extract meaningful insights from a text corpus.

**Table 3.** Research Questions (adapted from van Dinter et al. [40])

| No. | Research Questions (RQ) |
|---|---|
| RQ1 | Which phase of the SLR process is the focus of automation? |
| RQ2 | Which TM/AI techniques (Table A2 in Supplementary Material) have been employed for automation? |
| RQ2A | Which TM/AI models/algorithms have been explored for automation? |
| RQ2B | Which TM/AI models/algorithms were the most heavily explored? |
| RQ2C | What evaluation methodology and metrics are used? |
| RQ3 | How will the adoption of TM techniques facilitate SLRs? |

| **RQ4** | What is the improvement from employing TM techniques over a manual process? |
|---|---|
| **RQ5** | What are the open challenges and solution directions? |

### 3.2 Search Scope

The SLR scope needs to include the time frame for publication and the sources from where the articles are sourced. Based on the literature that we have seen so far a reasonable time frame was found to be 20 years so we decided to adhere to this. The year 2000 was kept as beginning of the search time frame and 2021 was fixed as the cutoff date. The language considered for inclusion was English and the reference type was journals, conference, workshops, symposiums and book chapters since so many existing SLRs restrict the reference type to these kinds of articles.

### 3.3 Search Method

We used automated search to retrieve relevant articles from publication repositories. Google Scholar was used as the primary retrieval tool and we also used snowballing in our search process to identify related relevant articles [41].

### 3.4 Search String

We tried various combinations of key words and Boolean operators to construct a search string. This was optimized iteratively to retrieve the maximum number of relevant results by manually conducting a number of trial searches. There is a character limitation of 256 chars in Google Scholar which required tuning the search string accordingly. This led to the following search query as the basis of the search:

("systematic literature review" OR "systematic review") AND ("Automation") AND ("Data Mining" OR "Text Mining" OR "NLP")

### 3.5 Criteria for Selection

Based on the research questions we have defined for this SLR we formulated the inclusion and exclusion criteria to fine tune the results from the search process. We followed a two phase screening process. Studies that satisfied the following inclusion criteria were included in the first phase:

- Publication year is after 2000
- The study describes a TM/AI/NLP process to support SLR automation (either complete automation or specific phases of SLR)
- If the SLR was a review of other SLRs then the reviewed articles are evaluated separately
- If multiple versions are available the most recent one is used

The output from the 1st stage screening process was then manually reviewed (2nd stage) with input from an external consultant to ensure that the results were relevant. Any record marked as doubtful, meaning that it is not clear if it is a relevant article, was discussed for a final decision regarding its inclusion.

The following exclusion criteria were applied to further fine tune the result set.

- The language of publication was not English
- Full text of the article was not accessible
- No empirical results were presented
- Not focused on automation of SLRs specifically
- Discusses TM/NLP/AI techniques but not in the context of SLR automation
- Focused on a particular commercial or open source tool (e.g. Abstrackr)

### 3.6 Quality Assessment

We assessed the quality of the studies in this stage that were being considered for inclusion. The papers in the search list were read in full text and a standard assessment criteria was applied to ascertain the quality score. We adapted the quality assessment criteria used by Feng et al.[14] which in turn was developed from the checklist provided by Dybå et al.[42] and Nguyen-Duc et al.[43] See table A1 in the Supplementary Material. Only articles scoring sufficiently high in quality, as detailed below, were included.

### 3.7 Data Extraction

We used a data extraction template to collect all necessary information from the selected primary list of literature to facilitate an in-depth analysis based on our research questions specified in Table 3. The main extraction elements used in that template are listed below. The detailed data extraction template is provided in Table A3 (Supplementary Material). We adapted the classification of TM methods as specified by Feng et al.[14] for use in the form for data extraction.

- Title
- Passed Inclusion Criter
- Year of Publication
- Authors
- SLR Steps Automated
- Level of Automation
- Type of Review
- TM Methods Used (Category)
- TM Model/Algorithm Information
- TM Model Evaluation Methodology Used (if Specified)
- Evaluation Metrics Used
- TM Methods Used as Additional Reviewer
- Deep Learning or AI Used?
- Sampling Techniques Used

- Overall Results/Conclusions (Stated by Authors)
- Performance Gain Over Manual Methods Provided?

### 3.8    Data Analysis

As part of the data analysis stage we analyzed the data extracted from the relevant arti-
cles from the earlier step, answering the research questions mentioned in Table 3. For
RQ1 we identified the specific phase of an SLR that is the focus of automation in the
article. For RQ2 we categorized the TM/AI techniques used by the researchers and
documented them appropriately using the TM categories in Table A2 (Supplementary
Material). In addition we also documented which techniques were found to be most
appropriate in the study and the evaluation metrics used in the process. For RQ3 and
RQ4 we gathered information on the adoption of TM techniques for SLRs and potential
performance gain observed over traditional methods. Finally for RQ5 we collected in-
formation on the open challenges that were observed either directly in an SLR or topics
that seemed inadequately addressed in the SLR. Scores were assigned based on quality
assessment criteria (Table A1 in Supplementary Material) as follows.

1. Quality assessment criteria were completely addressed (score was 3).
2. Quality assessment criteria were addressed with moderate gaps (score was 2).
3. Quality assessment criteria were addressed with considerable gaps (score was 1).
4. Quality assessment criteria were not addressed (score was 0).

The list of the papers was finalized after filtering on the quality assessment score as
explained below.

### 3.9    Data Synthesis

Data synthesis is the process of interpreting extracted data to answer the research ques-
tions of an SLR. Since each paper might have different naming standards for defining
the objectives, naming the algorithms used, models used, etc., we synthesized the data
collected using synonyms to gain information on the data patterns.

Using the methods described in the previous sections, this section collates and summa-
rizes the results.

**Search Results and Identification of Studies**. Using the search strategy mentioned in
the above sections we first searched the six electronic databases for relevant studies
using queries as described in section 3.4. The process to identify and finalize the set of
studies is depicted in Fig. 2. Table 4 shows the digital libraries that were searched and
the final number of papers that were shortlisted from each library after applying the
finalization process shown in Fig. 2.

For Google Scholar the initial set returned during the search using the criteria mentioned in section 3.4 was 15,500. Google Scholar would not provide records past 1,000. We observed that the relevancy of the search data set diminished rapidly after the first 600 records, and decided not to screen the remaining records based on this observation. The search results were then screened for relevancy ($1^{st}$ and $2^{nd}$ stage screenings). Full text documents were retrieved only for the screening final set, for detailed quality analysis and to determine the final list of articles to review.

The articles in the screening final set from Table 4 were then subjected to a quality review (QR), explained below to give the QR final set row. We then used the QR final set for a manual snowball (MS) search to ensure that more relevant results, if any, were added to the cumulative final list.

**Table 4.** Search Results for Each Digital Library

| Digital Library → | Google Scholar | Pub-Med | Web Of Science Core Collection | ACM | IEEE | Science Direct | Total |
|---|---|---|---|---|---|---|---|
| **Initial Set Returned** | 15,500 | 143 | 440 | 468 | 60 | 645 | 17,256 |
| **$1^{st}$ Stage Screening** | 63 | 19 | 63 | 77 | 13 | 12 | 247 |
| **$2^{nd}$ Stage Screening** | 53 | 19 | 60 | 32 | 11 | 8 | 183 |
| **Screening Final Set** | 28 | 3 | 27 | 14 | 1 | 4 | 77 |
| **QR Final Set** | 11 | 2 | 2 | 5 | 0 | 0 | 20 |
| **Initial MS Search** | 28 | 3 | 27 | 14 | 0 | 0 | 72 |
| **Screened MS Search** | 11 | 2 | 14 | 5 | 0 | 0 | 32 |
| **Finalized MS Search** | 11 | 2 | 14 | 5 | 0 | 0 | 32 |
| **Cumulative List** | 22 | 4 | 16 | 10 | 0 | 0 | 52 |
| **Final (No Duplicates) List** | 10 | 2 | 12 | 5 | 0 | 0 | 29 |

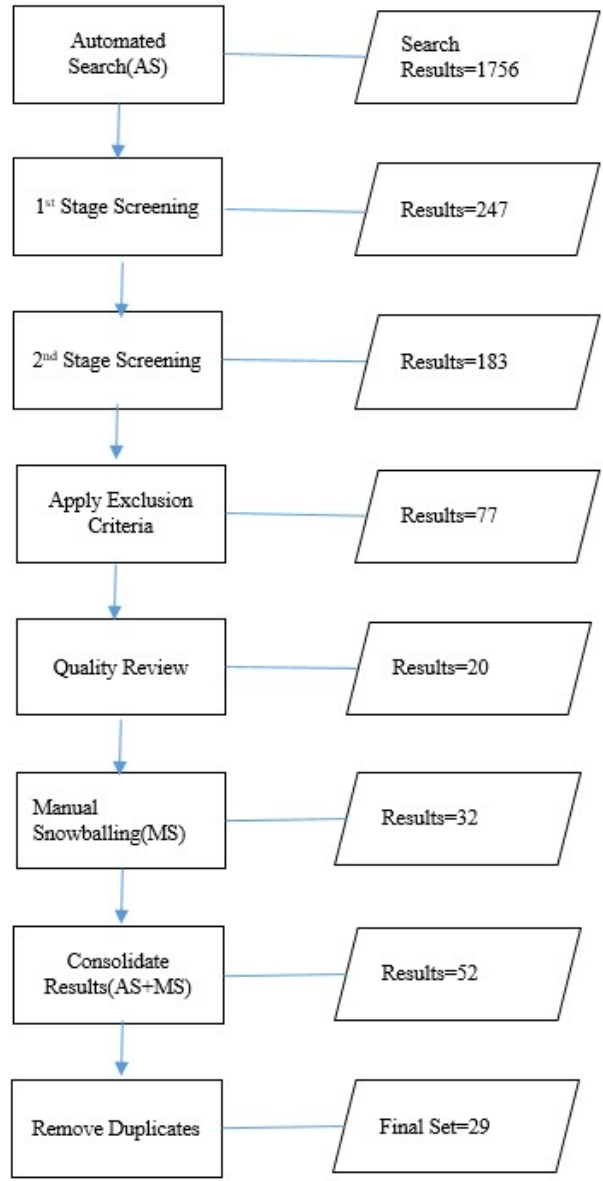

**Figure 2.** Selecting and Finalizing Studies

**Quality Assessment Results.** The QR final set shown in Table 4 was created after subjecting the 2nd stage screening results to a quality assessment review introduced in Section 3.6. Only articles having a quality assessment score >= 2 were included in the final list. Full texts for all of the articles were analyzed during this process. We followed the same procedure for the manual snowballing exercise as well. The Final List is the combination of the QR Final Set and Finalized MS Search rows in Table 4. This multi-step process of quality review and screening was designed to result in a final list of articles that were of high quality and the most relevant articles for our research questions.

## 4    Discussion and Conclusion

In this section we analyze the results and answer the research questions shown in Table 3. Table A4 in the Supplementary Material lists these studies for reference.

**RQ1**: Which phase of the SLR process is the focus of automation?

As part of our analysis, we identified 29 studies which were relevant based on the research questions and contribution to automation of the SLR steps mentioned in Table 1. Some studies were focused on automating multiple stages. During our analysis we derived the following insights.

- 24 studies focused on automating stage SLR6, selection of studies.
- 8 studies focused on automating stage SLR8, data extraction and monitoring.
- 1 study focused on stage SLR9, data synthesis.
- 1 study focused on stage SLR5, identification of research.

**RQ2:** Which categories of TM/AI techniques (Table A2, Supplementary Material) have been employed for automation?

Twenty-four studies (77%) were related to classification (categorization), two (6%) were related to clustering, two to information extraction (IE), two to information retrieval (IR) and one (3%) to summarization. Clearly the preponderance of the studies used TM/AI methods for classification.

**RQ2A:** Which TM/AI models/algorithms have been explored for automation?

We analyzed the data extracted from the finalized set of studies which is documented in Table A4 (Supplementary Material).

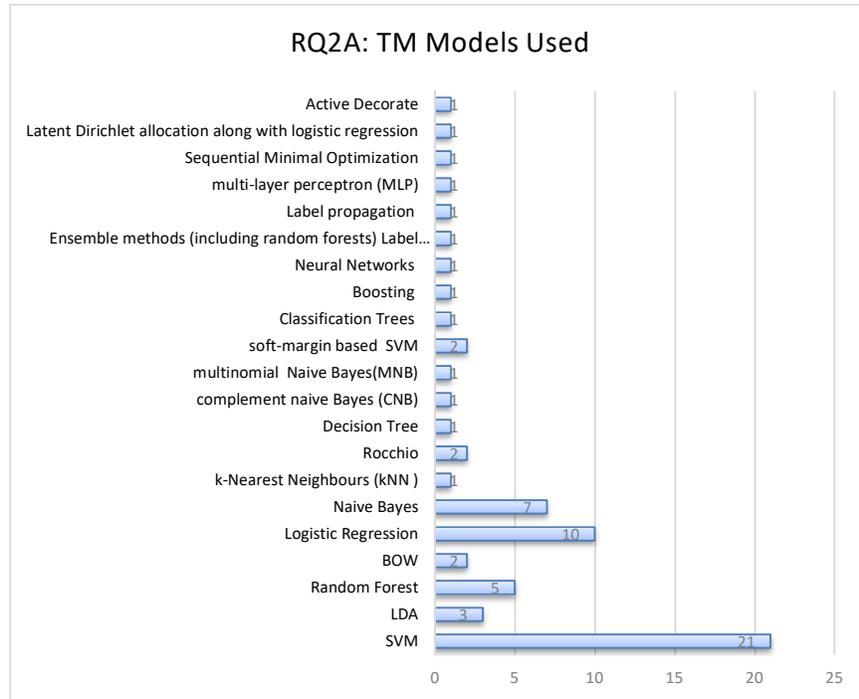

**Figure 3.** Most Frequently Used TM Models for Experimentations

**RQ2B:** Which TM/AI models/algorithms were the most heavily explored?

As indicated in Fig. 3, SVM, logistic regression, naïve Bayes, and random forests are the most frequently used algorithms.

**RQ2C:** Which evaluation methodologies and metrics have been used?

An evaluation metric as "A metric that quantifies the performance of a predictive model [44]. This typically involves training a model on a dataset, using the model to make predictions on a holdout dataset not used during training, then comparing the predictions to the expected values in the holdout dataset." We found that cross validation was the most frequently used evaluation methodology.

**RQ3:** How will the adoption of TM techniques facilitate SLRs?

As Marshall et al. [16] mention in their paper the most frequently used application of TM/NLP techniques in the SLR field is text classification and data extraction. We arrived at the same conclusion as part of our SLR as well. Classification methods are generally used to classify the paper in question as relevant or not based on the research questions the SLR is attempting to address. Data extraction on the other hand is used to extract important portions from the SLR to get data for a particular variable or attribute of interest. For example extracting the PICO elements from an SLR is a very common application of data extraction. TM techniques appear to have the potential to significantly impact the quality and speed of SLR development.

**RQ4:** What is the improvement from employing TM techniques over manual processing?

There was mention of improvement due to using the NLP/TM model. Wallace et al. [45] mention that they were able to reduce the number of citations to be screened by 40-50% without excluding any relevant ones. Pham et al. [46] achieved workload reduction in the range of 55-63% with the number of missed studies in the range of 0-1.5%. Norman et al. [47] found that the main meta-analysis for each systematic review can be reliably performed with an estimation error of 1.3% average after screening around 30% of the candidate articles.

**RQ5:** What are the open challenges and solution directions?

As mentioned in explaining RQ1 above, the majority of the research has been focused on SLR6 (selection of sstudies), and a distant second is SLR8 (data extraction and monitoring). Important SLR activities such as SLR5 (identification of research, including search string development and literature search), SLR7 (assessing study quality) and SLR9 (data synthesis) have been less scrutinized. We also found that there is need and opportunity for continued development of artificial intelligence techniques in the SLR process.

## 4.1    Summary of Conclusions

In this SLR we have described the use of NLP/TM techniques in the area of automation of SLR development. This area has been active for the last decade and continues to attract more research. As mentioned in RQ5 there are significant gaps in the application of TM and other automation techniques in the areas of data extraction, monitoring, quality assessment and data synthesis. AI in the SLR automation process has experienced a recent surge of exploration and there is a need to continue this due to the promise of improved automation and all the benefits flowing therefrom.

## 4.2   Future and Ongoing Research

The primary objective of conducting this systematic literature review was to find the current state of the art in the field regarding applying NLP, TM and AI techniques to automate specific steps in an SLR. After conducting this SLR we found that there are significant opportunities to use NLP to assist SLR and more specifically in the data extraction phase. We are currently conducting research on using various NLP techniques to assist in extraction of PICO [22] data elements from randomized control trial free text articles and summarizing the overall clinical evidence.

## Supplementary Material

The supplementary material, including data tables A1-A4, may be obtained at: https://dberleant.github.io/papers/sundber-supp.pdf.

## Acknowledgments

Publication of this work was supported by the National Science Foundation under Award No. OIA-1946391. The content reflects the views of the authors and not necessarily the NSF. The authors are grateful to Deepak Sagaram, MD, for consulting on the list of articles regarding their relevance for inclusion and exclusion.